# Challenges in Aligning Requirements Engineering and Verification in a Large-Scale Industrial Context

Giedre Sabaliauskaite[1], Annabella Loconsole[1], Emelie Engström[1], Michael Unterkalmsteiner[2], Björn Regnell[1], Per Runeson[1], Tony Gorschek[2], Robert Feldt[2]

[1] Department of Computer Science, Lund University, Sweden
[2] Blekinge Institute of Technology, Sweden
{Giedre.Sabaliauskaite, Annabella.Loconsole, Emelie.Engström,
Björn.Regnell, Per.Runeson}@cs.lth.se
{Michael.Unterkalmsteiner, Tony.Gorschek, Robert.Feldt}@bth.se

**Abstract.** [**Context and motivation**] When developing software, coordination between different organizational units is essential in order to develop a good quality product, on time and within budget. Particularly, the synchronization between requirements and verification processes is crucial in order to assure that the developed software product satisfies customer requirements. [**Question/problem**] Our research question is: what are the current challenges in aligning the requirements and verification processes? [**Principal ideas/results**] We conducted an interview study at a large software development company. This paper presents preliminary findings of these interviews that identify key challenges in aligning requirements and verification processes. [**Contribution**] The result of this study includes a range of challenges faced by the studied organization grouped into the categories: organization and processes, people, tools, requirements process, testing process, change management, traceability, and measurement. The findings of this study can be used by practitioners as a basis for investigating alignment in their organizations, and by scientists in developing approaches for more efficient and effective management of the alignment between requirements and verification.

## 1 Introduction

Are we sure that the tests performed are based on requirements and not on technical specifications supplied by developers? Are we sure that the test coverage is adequate? In order to assure that customer requirements are realized as intended these questions must be asked and answered. However, this is not an easy task, since requirements tend to change over time [13], and in many cases the requirement specifications are not updated during the development of a product making it hard to use them as a solid base for creating e.g. test cases [7, 15]. In small systems with just a few requirements it could still be possible to handle the changes manually, but it gets extremely hard in complex systems with thousands of requirements. Therefore, there is a need for a mechanism to manage coordination between the requirements and the verification processes. We call such coordination *alignment*.

In this paper, we examine the challenges in aligning the requirements and the verification processes. We present preliminary results of an interview study performed in a large software company in Sweden. The overall goal of our research is to understand how alignment activities are performed in practice, what the important problems are and what can be improved to gain better alignment. The results presented in this paper are a set of challenges that can help practitioners and researchers. Practitioners can, for instance, allocate more resources in the areas that are challenging when aligning the requirements and the verification processes. Researchers can also benefit from our results by focusing their research on the areas that are the most challenging. The results are valid in the context of the company under investigation. We are currently extending the case study to other companies. By comparing the results of this case with other case studies it will be possible to get a more general picture of challenges in different kinds of organizations and in different domains.

The paper is structured as follows. In Section 2, we present related work in the area. In Section 3, we describe the research approach used in this qualitative case study. Section 4 describes the alignment challenges found. Finally conclusions and future work are presented in Section 5.

## 2 Related Work

In [17], authors presented the findings of the discussions with test managers and engineers in software development organizations regarding difficulties of integrating independent test agencies into software development practices. The company where we have performed interviews does not commonly use independent test agencies, however it has separate requirements, development, and testing units. Therefore it would be interesting to compare the results of having independent test agency and independent test unit within the company under study.

Findings related to change management emphasize the importance of synchronization between the development and test with respect to modifications of functionality [17]. The results of our study confirm these findings. One of the most recurrent challenges identified in our study is that requirements are not being updated on time.

Findings related to people interactions and communication stress the need of communication between development and test organizations. If testers do not know who wrote or modified the code, they do not know whom to talk to when potential faults are detected. On the other hand, it could be difficult for developers to inform testers on upcoming functionality changes, if they don't know whose test cases will be affected [17]. Our study confirms these results as well. Most of the interviewees suggest that alignment could be greatly improved if requirements and testing people would interact more with each other.

Several surveys on requirements related challenges are present in the literature: problems in the requirements engineering process [9], requirements modeling [6], quality requirements [7], requirements prioritization [1], and requirements interdependencies [8]. Among these, Karlsson et al. [1] have results similar to ours, i.e. tool integration is difficult and it is a challenge to write quality requirements.

Most of the studies above do not focus on the alignment between the requirements and the verification processes. Research in connecting requirements and testing has been performed by several authors, for instance Uusitalo et al. [4], Post et al. [3], and Damian and Chisan [10]. Uusitalo et al [4], have conducted a series of interviews in order to investigate best practices in linking requirements and testing. Among the best practices, authors mention early tester involvement in requirements activities. They conclude by suggesting to strengthening the links between requirements engineers and testers, since it is difficult to implement traceability between them; a conclusion supported by this study (see Section 4.7).

The importance of linking requirements and verification is also stressed by Post et al. [3]. They describe a case study showing that formalizing requirements in scenarios make it easier to trace them to test sets. Damian and Chisan [10] present a case study where they introduce a new requirements engineering process in a software company. Among the practices in the process, they include traceability links between requirements and testing, cross-functional teams, and testing according to requirements. They show that an effective requirements engineering process has positive influence on several other processes including testing process.

The case studies above [3, 4, 10] are performed in a medium scale requirements engineering context [11], while our study is performed in a large/very large scale context and includes many aspects of aligning requirements and verification.

## 3  Research Approach

The approach used in this study is qualitative. Qualitative research consists of an application of various methods of collecting information, mainly interviews and focus groups. This type of research is exploratory [16]. Participants are asked to respond to general questions, and the interviewers explore their responses to identify and define peoples' perceptions and opinions about the topic being discussed. As the study was meant to be deep and exploratory, interviews were the best tool since surveys are not exploratory in nature. The interviews were semi-structured to allow in-depth, exploratory freedom to investigate non-premeditated aspects.

In this study, we interviewed 11 professionals in a large software development company in Sweden, based on the research question: What are the current challenges in aligning the requirements and the verification processes?

The viewpoint taken in this research is from a process perspective. The researchers involved do not work directly with artifacts, but with processes and have expertise in fields like requirements, testing, quality, and measurement.

Based on our pre-understanding of the processes involved in aligning requirements and verification, a conceptual model has been designed (see Figure 1). This model was used as a guide during the interviews. In this model, we consider three dimensions of requirements and test artifacts, connected through work processes. One is the *Abstraction level dimension*, from general goals down to source code, which is similar both for the requirements and the testing side. Test artifacts are used to verify the code, but also for verifying the requirements. The arrows are relationships that can be both explicit and implicit, and can be both bi- or uni-directional. Then, we have the *Time dimension*, in which the processes, the products, and the projects change and evolve. This has an effect on the artifacts. There is also the *dimension of Product lines*, which addresses variability, especially applicable when the development is based on a product line engineering approach [2].

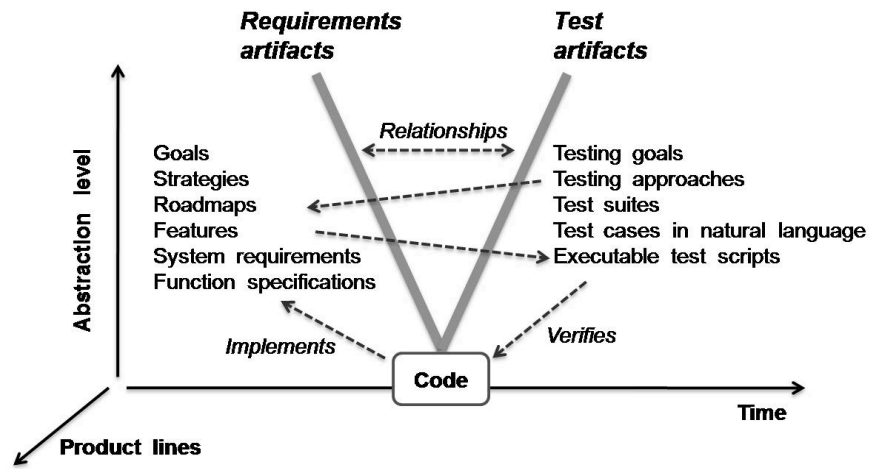

**Fig. 1**. Conceptual Model

**Case Context**. Our results are based on empirical data collected through interviews at a large anonymous company, which is using a product-line approach. The company is developing embedded systems for a global market, and has more than 5000 employees. A typical project in this company lasts about 2-years, involves circa 800-1000 men per year, and has around 14000 requirements and 200000 test cases. The tool DOORS is used for requirements management, and the tool Quality Center for test management. Further information about the company is not disclosed for confidentiality reasons.

The interviews have been distributed in time between May and October 2009.

### 3.1 Research Methodology

In this study, challenges and problems, as well as, current good practices and improvement suggestions regarding alignment between the requirements and verification processes have been identified through interviews with software engineering practitioners. The results from 11 interviews are included in this paper. Employees with different roles have been interviewed: quality management related roles (quality manager and quality control leader), requirements related roles (requirements process manager, requirements architect and requirements coordinator),

developer and testing related roles (test leader, tester). The research was conducted in several steps:

1. Definition of interview guide;
2. Interview planning and execution;
3. Transcription of interviews, and division of transcriptions into sections;
4. Definition of codes (keywords) to be assigned to transcriptions' sections;
5. Coding of interview transcriptions using predefined codes;
6. Sorting of coded transcriptions to group transcription sections according to codes;
7. Analysis of the results;
8. Validation of results by feedback to the organization.

**Step 1.** We constructed an interview guide, which is a document containing 30 questions to be asked during the interviews. The first version of the interview guide contained 22 questions, defined based on the research questions. The questions were validated during 2 pilot interviews. This led to the updated list of questions, grouped into several topics, such as general questions about requirements and testing, questions on quality requirements, etc. An overview of the interview guide is available in Table 1[1].

**Table 1.** Overview of the interview guide

| Interview topics | Description |
|---|---|
| Software requirements | Handling of functional and quality requirements, customer involvement |
| Software testing | Handling of testing artifacts, customer involvement, testing of functional and quality requirements |
| Alignment between requirements & verification processes | Alignment importance, current alignment method (documents, processes, methods, tools, principles, practices, etc.), alignment responsible, problems & challenges, improvement ideas & expected benefits |
| Measurements and feedback gathering | Alignment related measurements, performance indicators, customer satisfaction evaluation |
| Product line engineering | Handling of requirements and testing, maintaining alignment |
| Outsourcing | Maintaining alignment in case of outsourcing |

**Step 2.** Eleven professionals were interviewed; each interview lasted for about one and a half hour. All interviews were recorded in audio format and notes were taken. A semi-structured interview strategy [16] has been used in all interviews, where the interview guide acted as a checklist to make sure that all important topics were covered. 2-3 interviewers interviewed one interviewee. One of the interviewers lead the interview, while the others followed the interview guide, took notes, and asked additional questions. The selection of the interviewees has been made based on recommendations by requirements managers, test managers, and the interviewees themselves. (At the end of each interview we asked the interviewees if they could recommend a person or a role in a company whom we could interview in order to get alignment related information).

---
[1] The complete version of the interview guide and coding guide are available at:
http://serg.cs.lth.se/research/experiment_packages/interview_study_on_requirements_verification_alignment/

**Step 3.** Interviews were transcribed into text in order to facilitate the analysis. The transcriptions were then divided into text sections containing 1-2 sentences. All the text sections have been numbered in order to keep the order of the sentences. The size of the transcriptions ranged from 4000 words to about 9000 words per interview.

**Step 4.** As suggested by C.B. Seaman [12], codes (keywords) were assigned to the transcriptions' sections in order to be able to extract all the sections related to a specific topic. However, the definition of the coding scheme turned out to be a non-trivial task. We started by making an initial list of possible codes, which included codes related to our research questions, alignment methods, quality requirements [14] and software development process activities. In order to extend and tailor this initial list of codes to our interview context, we decided to perform exploratory coding [16], which included six researchers analyzing several interview transcriptions individually and assigning suitable codes to the text sections.

The result of exploratory coding was a list with 169 codes. In the next stage, we reviewed the codes resulting from the exploratory coding, grouped them into several categories at different abstraction levels and developed a coding guide. The coding guide is a document containing the list of codes and detailed instructions of how to code a transcription. In order to validate the coding guide, seven researchers used it to code the same interview transcription (let's call it X) individually, and then had a meeting to discuss differences in coding and possible improvements of the coding guide. Kappa inter-rater agreement [18] has been used as a metric to evaluate improvement in homogeneity of coding by different researchers. Consequently, the coding guide was updated and the interview transcription (X) was coded again using the updated version of the coding guide to make sure that the differences between different coders were minimized. The coding guide included codes at three abstraction levels: high, medium, and low (see Table 2). The high-level codes were based on research questions. The medium-level codes included different categories relevant to our research, and the low-level codes were the coder's interpretation of the transcription's section. A summary of the codes is presented in Table 2.

**Table 2.** Overview of the codes assigned to transcription's sections (see footnote 1 for a complete list of codes).

| Abstraction level | Description |
|---|---|
| High | Codes related to research questions, i.e. alignment practices, problems and challenges, improvement ideas and benefits |
| Medium | Two groups of codes: Group 1 – thirteen categories, which include requirements, testing, traceability, configuration management, organization processes, interactions, product quality aspects, and measurements among others. Group 2 – additional categories, e.g. product-line engineering, outsourcing, open source. |
| Low | Coder's interpretation of the transcription's section, a brief summary of the information described in the section |

**Step 5.** Eleven interview transcriptions were randomly assigned to four researchers, who coded them using the final version of the coding guide. The template used during coding is shown in Table 3.

**Table 3.** Template used during coding.

| No | Text | High-Level Coding | Medium-Level Coding | | | Low-Level Coding, Comments |
|----|------|-------------------|---------------------|---------|---------|---------------------------|
|    |      | Research Questions | Group 1 | | Group 2 | |
|    |      |                   | Primary | Secondary | | |
|    |      |                   |         |           |         | |

**Step 6.** Coded interview transcriptions were merged into one file, making it possible to group transcription sections according to codes.

**Step 7.** The identified transcription's sections of each group were analyzed by two researchers. In order to identify alignment challenges, researchers studied all the transcription's section coded as "challenges" with the goal to extract challenges from the information provided by interviewees. Some challenges were similar and therefore could be reformulated or merged together, while others were kept apart as they were different.

**Step 8.** The results of the analysis were validated by feedback from the organization where the interviews have been conducted.

### 3.2 Validity Discussion

A discussion of possible threats to validity will help us to qualify the results and highlight some of the issues associated with our study. As suggested by P. Runeson and M. Höst [5], we have analyzed the construct validity, external validity, and reliability. Internal validity is concerned with threats to conclusions about cause and effect relationships, which is not an objective of this study. A detailed list of possible threats is presented in [16].

**Threats to Construct Validity.** The construct validity is the degree to which the variables are accurately measured by the measurement instruments used in the study [5]. The main construct validity threat in this study regards the design of the measurement instrument: are the questions formulated so that the interviews answer our research questions? Our main measurement instrument is the interview guide (see Section 3.1, Step 1), which includes the questions to be asked. Two researchers have constructed it by analysing the research questions and creating sub-questions. Five researchers have reviewed it to check for completeness and consistency; therefore we believe that the interview guide is accurate. The other measurement instrument is the coding guide. As described in Section 3.1, Step 4, this instrument has been validated by seven researchers in order to make sure that the result of the coding activity had minimal individual variations.

The questions in the interview guide were tailored on the fly to the interviewees since the professionals participating in the interviews had different roles and different background. Our study is qualitative; the goal is not to quantify answers of the same type, rather to explore the different activities in the company, which could be done best by investigating deeply the role of each interviewee.

Another potential threat in this study is that different interviewees may interpret the term "alignment" differently. For this reason, the conceptual model (see Figure 1) has been shown to the subjects during the interviews, in order to present our definition of alignment between requirements and verification.

**Threats to External Validity.** The threats to external validity concern generalisation. The purpose of this study is not to do any statistical generalization of the results to other contexts, but to explore the problems and benefits of alignment in the context of the specific company. The study was performed in an industrial environment where the processes were real, and the subjects were professionals. Hence, we believe that the results can be analytically generalized to any company of similar size and application domain. The company might not be representative; therefore more companies will be interviewed in order to get results independent of the kind of company.

**Reliability.** Reliability issues concern to what extent the data and the analysis are dependent on the researchers. Hypothetically, if another researcher later on conducts the same study the results should be the same. In this study, all finding have been derived by at least two researchers, and then reviewed by at least three other researchers. Therefore, this threat has been made smaller.

In our study, the investigation procedures are systematic and well documented (see Section 3). The interview guide, the researchers' view (the conceptual model), and the coding scheme were reviewed independently by seven researchers with different background.

The presented observations reflect the views of the participants. The interviews have been recorded and transcribed. The transcriptions could contain errors due to misinterpretation, mishearing, inaccurate punctuation or mistyped words. In order to minimize these threats, the transcriber has also been present at the interview. Moreover, the transcriptions were sent to the interviewees so that they could correct possible misinterpretation of their answers.

One factor affecting the reliability of the data collected can be the fact that the interviews capture the subjective opinion of each interviewee. However, we interviewed 11 professionals, which we believe is a sufficient amount to capture the general view of the company. Influence among the subjects could not be controlled and we could only trust the answers received. The choice of the subjects in the company might not give a representative picture of the company; however, the subjects had different roles and we tried to cover diverse roles within the company.

Regarding the coding activity, it is a classification of pieces of text, which are taken out of context; hence there is a risk of misinterpretation. This risk was minimized by checking the whole context of the text while doing data analysis.

To summarize, we believe that the validity threats of our results are under control, although the results should not be generalized to all organizations.

## 4  Analysis and Result

The result of this study includes a range of challenges faced by the studied company grouped into these categories: organization and processes, people, tools, requirements process, testing process, change management, traceability, and measurement. The grouping is rough: if the challenge belonged to several categories, we assigned it to the category which was the most relevant. The choice of the categories was based on the medium level codes (see Table 2). All challenges are rooted in the interview transcriptions. The challenges of each group are presented in Subsections 4.1-4.8.

### 4.1  Organization and Processes Related Issues

This section summarizes the alignment problems and challenges related to the company's organizational structure and processes.
- The requirements and verification processes are separate processes and are not aligned. Furthermore, processes can use different standards of documentation, which negatively influence the hand-over between different parts of organization. Moreover, some parts of the company follow a documented development process while other parts do not.
- Frequent process changes negatively influence alignment. It would take time for people to learn and use the new process. Sometimes, people are reluctant to use a process knowing that it will change soon. Also, some good practices could be lost due to the process changes.
- Distance in time between the development of requirements and test artifacts can create alignment problems. Requirements can be approved without having test cases associated with them. This can result in having non-testable requirements.
- In a large company, gaps in communication across different organizational units often occur, especially at the high level. Furthermore, as stated by an employee "*it is hard to find who is accountable for things because depending on who you ask you get very different answers*". Therefore, this could affect the alignment, especially at the high abstraction level of the requirements and verification processes.
- Implementation of process improvements is time consuming, especially when the improvements are involving several units. Several issues related to the management can affect the alignment, e.g. decisions are not documented, lessons learnt are not always collected and processes depends on individual commitment.

Summarizing the challenges, the requirements and the verification processes are not aligned and are distant in time. There are also communication problems across different organizational units and the decisions are not documented, therefore it is hard to know who is accountable for a decision. The organizational structure and the processes, as well as changes in these are influencing the alignment. One reason could be that the company is very large and many organizational units are involved, and not every unit follows the documented process, and the standard for documentation.

### 4.2 People Related Issues

This subsection presents a list of issues that are related to people, their skills and communication with each other.
- Professionals do not always have good technical knowledge and understanding about the work of other units. Requirements engineers sometimes lack knowledge about implementation as well as testing, while testers lack knowledge of requirements. Also, professionals are sometimes unwilling to move within the company in order to gain this knowledge. This has a negative effect on alignment between requirements and verification processes.
- Lack of cooperation between requirements people, developers and testers is affecting the alignment. In some cases, requirements engineers and developers have a good communication, as well as developers and testers. However, when there is a lack of direct communication between requirements and testing people, alignment is influenced negatively.

The main challenge in this area is communication, cooperation, and understanding of each other's work within the company. This can be hard when working under tight deadlines; there is no time to communicate and understand each other's work. Adequate technical knowledge, communication and cooperation between requirements people, developers and testers greatly influence the alignment.

### 4.3 Tools Issues

Software tools play a crucial role in maintaining alignment between different artifacts. The following are several tool related issues.
- The lack of appropriate tools influences the alignment. It is very important to have reliable and easy to use requirements and verification tools. If the tool is difficult to use, or it is not reliable, people are not willing to use them. Having a good requirements management tool, which includes not only information about requirements, but also the flow of requirements, is crucial for testers. Otherwise, testers try to get this kind of information from other sources, for instance the developers. Tools for managing quality requirements are needed, otherwise there is a risk that quality requirements are not implemented and/or tested.
- It is important to keep the requirements database updated. If requirements are not up to date, testers will test according to old requirements and will find issues, which are not really failures, but valid features.

- If there is no tool to collect customer needs, it is difficult to keep them aligned with requirements, hence with test cases as well. And this leads to misalignment between customer needs and requirements, and consequently affects customer satisfaction with the final product.
- In cases when requirements and testing artifacts are stored in different tools, there is a need of good interfaces between these tools, and access of all interested parties to the tools. Otherwise, it becomes very difficult to maintain alignment. Especially when there are many-to-many relationships between requirements and test cases.
- If the mapping between requirements and test cases is not presented in a clear way, it could contain too much redundant information, and therefore it could be difficult for requirements people and testers to use it.

Most of the interviewees stated the lack of adequate software tools, which would allow to handle requirements, verification, and to measure the alignment between them. Furthermore, the interface of the tools and tool integration is not always good. The consequence of this is that people become reluctant to use them and do not update the information stored in them. This is greatly affecting the alignment.

**4.4  Requirements Process Related Issues**

This subsection presents a list of issues that are related to the requirements process.
- Requirements sometimes are not given enough attention and consideration by other organizational units, such as development and testing units. According to an employee "*Developers do not always review the requirements, and discover requirements that can not be implemented during development, even when having agreed on the requirements beforehand*". This could be due to the lack of involvement of developers and testers in requirements reviews.
- Not having a good way of managing customers' needs makes it more difficult to define requirements, especially requirements at a high abstraction level.
- Requirements engineers do not think about testability of requirements. Therefore, requirements could turn out to be non-testable.
- Dealing with quality requirements is a difficult task. Quality requirements tend to be badly structured or vague. Furthermore, it is difficult to assign quality requirements to different development groups for implementation, since several groups are usually involved in implementing a quality requirement, and none wants to take a full responsibility for that.
- It is difficult to maintain alignment in organizations working with a large set of requirements, when the number of requirements reaches tens of thousands or more. Furthermore, in the organizations, which are using a product lines engineering [2] approach, maintaining alignment between domain and application requirements and test cases could be a challenge.

As we can see, there are numerous challenges related to requirements process, which affect alignment. Most of the interviewees stress the importance of updating requirements as soon as changes occur, and finding adequate ways of defining and managing quality requirements. These two are the most recurrent requirements process related challenges.

### 4.5 Testing Process Related Issues

The following are the issues that related to the testing process.
- Sometimes testers lack clear directions on how to proceed with testing. Especially while testing high-level requirements, such as roadmaps for example. It is difficult to test that the products adhere to roadmaps, since such testing takes a long time and is costly. Usually short loops are preferred.
- In case several organizational units are involved in testing, the cooperation between them is crucial. It is particularly relevant to the companies, which have a product line engineering approach, since different organizational units could be performing domain and application testing, and the faults detected in applications should be removed from domain as well.
- There is a lack of verification at early development stages, especially of quality requirements verification. This results in lower quality of the product, as well as added cost and time spent on removal of defects at later development phases.
- It is inefficient to maintain alignment of requirements and test cases due to the large amount of test cases; sometimes their number reaches hundreds of thousands.
- It is difficult to get requirements people interested in having good quality test cases. Requirement people's involvement in reviewing test cases contributes to alignment, since this would help to assure that test cases comply with requirements.

As we can see from the above-mentioned challenges, there are numerous testing process related issues that can affect alignment. Having a well defined testing process at different development stages, and good cooperation between testing units could help improve the alignment.

### 4.6 Change Management Issues

The following are the challenges related to change management.
- It is sometimes difficult to find the people responsible, if a change occurs, if a defect is found, or if there is a need of further information. Thus, requirements engineers do not always inform related developers and testers in case of a requirements change. Furthermore, if a failure is found during maintenance phase, it is extremely difficult for maintenance people to find requirements people who can give information regarding requirements, or whom to inform about implemented changes. Therefore, maintenance people sometimes need to use testers as a source of information about requirements.
- There is a lack of strategy in deciding which changes to implement in case there is not enough time or resources to implement all changes.

- The information about changes is not always timely updated in the requirements database. Therefore it is difficult for developers and testers to know that the change has occurred.

Updating the requirements on time is one of the most recurrent challenges. It is therefore important to find ways to cope with changes immediately so that the traceability with testing can be maintained. In addition, delta handling and good tracking and reporting on the requirements and test case tools is needed to easily track changes and verify completeness.

### 4.7 Traceability Issues

The following are the challenges related to traceability between requirements and testing artifacts.
- There is a lack of links between requirements and test cases. Some test cases are very complex; therefore it is difficult to trace them back to requirements.
- If traceability between requirements and test cases is not maintained, testers keep testing requirements that have been removed. The reasons for lack of traceability could be the difficulty to implement traceability, and the lack of resources to maintain it.
- Having large legacies implies that a lot of test cases do not have requirements linked. This complicates implementation of alignment.
- Ideally, alignment should be implemented and maintained at all abstraction levels of requirements and verification processes. However, if it cannot be done for various reasons, such as lack of resources or time constraints, it is necessary to clearly define at which level to implement alignment.

The main challenge is the large volumes and complexity of requirements, test cases and test results. These are negatively influencing traceability. Better tools could help in managing the traceability in large scale requirements engineering and testing.

### 4.8 Measurements Issues

The following are the measurements related challenges.
- Due to the lack of experience in using measurements, it is difficult to define appropriate metrics or indicators.
- There is a lack of alignment related metrics. For example, one of the alignment metrics is requirements coverage by test cases, which is measured by calculating a percentage of requirements that have associated test cases. However, if a requirement has several test cases associated to it, it still could lack complete test coverage. Therefore, additional metrics are needed in order to get more complete information about requirements coverage.
- Key Performance Indicators (KPI) and metrics should be appropriate at both operative, as well as, top management level. Sometimes KPIs are useful only at top management level, but do not provide important information at the operative level regarding the things that could be improved.
- Sometimes target values for metrics and indicators are defined without a business case, not based on historical measurement data. Therefore, they could be inachievable.

Among challenges regarding measuring the alignment that are mentioned, the most recurrent is the difficulty of defining metrics to measure the alignment, especially the requirements coverage. Definition and use of adequate alignment metrics could help improve the alignment.

## 5  Conclusions and Further Research

In this paper, we have presented results of an interview study performed in a large software development company in Sweden. The goal of the study was to explore the current challenges in aligning requirements and verification processes.

One of the main challenges found regards software tools, both for managing requirements and for managing test cases. Often tools are not easy to use, and when different tools are used, the interface between them is poor. The consequence is that employees tend to not update the requirements stored in the tools and the information stored becomes obsolete and not useful. Traceability is also a challenge, and its importance is corroborated by other studies [3, 14]. Communication and cooperation across different units within the company is also a major challenge, confirming the results in [1, 17]. As a consequence of the challenges, company has decided to improve it's development process.

Our results can inspire other practitioners in their alignment improvement efforts since they can learn from this case what can be the most salient challenges in managing large quantities of requirements and test information in natural language. Researchers can also learn from this study since they can focus their research on existing challenges of potentially general interest.

We are extending this study to other companies of different size and domain. This will further enhance a general picture of alignment issues.

**Acknowledgements.** This work was funded by the Industrial Excellence Center EASE - Embedded Applications Software Engineering, (http://ease.cs.lth.se). Many thanks to the anonymous interviewees for their dedicated participation in this study, and reviewers of the paper for their valuable comments.